\documentstyle[twoside,fleqn,espcrc2]{article}
\title{
\vspace{-35mm}
\rightline{\small DESY 01-149~}
\rightline{\small HU-EP-01/43}
\vspace{10mm}
Performance studies of the two-step multiboson
algorithm in compact lattice QED
\thanks{Talk presented by N.V.~Zverev at LATTICE 2001, Berlin.} 
}
\author{I.L.~Bogolubsky$^{\rm a}$, V.K.~Mitrjushkin%
\address{Joint Institute for Nuclear Research, 141980 Dubna, Russia},
I.~Montvay%
\address{Deutsches Elektronen-Synchrotron, 
22603 Hamburg, Germany},
M.~M\"uller-Preussker$^{\rm c}$ and N.V.~Zverev%
\address{Humboldt-Universit\"at zu Berlin, Institut f\"ur Physik,
10115 Berlin, Germany}
}
\begin{document}

\begin{abstract}
The performance of the two-step multiboson (TSMB) algorithm
is investigated in comparison with the hybrid Monte Carlo (HMC)
method for compact lattice QED with standard Wilson fermions
both in the Coulomb and confinement phases. The restriction to
QED allows for extensive measurements of autocorrelation times.
Preliminary results show that the TSMB algorithm is at least
competitive with standard HMC.
\vspace{1pc}
\end{abstract}

\maketitle

Compact lattice 4D QED with $N_f=2$ dynamical Wilson fermions has been 
numerically studied \cite{HfMiMPNhSt} by means of the HMC 
algorithm \cite{DuKePeRow}. However, the use of this algorithm
within the confinement phase in the chiral limit 
$\kappa \rightarrow \kappa_c(\beta)$ 
as well as in the strong coupling phase with $\kappa > \kappa_c$
(the so-called 'Aoki phase') becomes cumbersome \cite{HfMiMPSt} 
because of large average condition numbers
\begin{equation}\label{cond_num}
\zeta=\frac{\langle\lambda_{\rm max}\rangle}%
{\langle\lambda_{\rm min}\rangle}.
\end{equation}
Here $\langle\lambda_{\rm max}\rangle$ and
$\langle\lambda_{\rm min}\rangle$, respectively, are the largest and
smallest average eigenvalues of the even-odd preconditioned
Wilson fermion matrix ${\cal M}^\dag{\cal M}$. Another open principal
problem for the HMC method is the simulation of an odd number of
fermion flavours. Hence, an alternative to the HMC method is desired.
The same arguments hold also for lattice QCD and related models.

An alternative is provided by the hermitian two-step multiboson (TSMB)
algorithm invented recently by one of the present authors (I.M.) 
\cite{Montv1}. It consists of the following basic ingredients. First, 
a polynomial approximation of the inverse fermion determinant
\cite{Lusch} is introduced. Second, in order to reduce the
computational costs \cite{Pear,BorFor,Jans,Bor} 
a noisy correction accept-reject step is employed.
At the end one performs an auxiliary reweighting
step for correcting the measured observables.

The TSMB algorithm has been successfully applied to a 
supersymmetric model \cite{Montv1}. Here we would like to investigate 
the efficiency of this algorithm in comparison with the standard HMC one.
For this aim compact lattice QED with Wilson fermions provides a valuable  
test ground, because its fermionic properties
- in particular in the confinement phase close to the chiral limit -
resemble very much those of lattice QCD with Wilson fermions  
\cite{HfMiMPSt}. In this paper we consider always the case $N_f=2$.

First of all, we want to establish a strategy for appropriately choosing
the technical parameters for both the HMC and the TSMB algorithms.
In order to achieve a reasonable acceptance rate for HMC we adjust
the number of time steps $N_\tau$ and the time step size
$\Delta\tau$ in the Hamilton dynamics according to the rule
\cite{EdwHrvKe}
\begin{equation}\label{dt_nhmc}
N_\tau \propto(\zeta^{1/4}\Delta\tau)^{-1},
\qquad \Delta\tau\propto (V\zeta)^{-1/4},
\end{equation}
where $\zeta$ is defined in (\ref{cond_num}), and
$V=N^3_s\times N_4$ denotes the lattice volume. The stopping criteria
for the conjugate gradient (CG) invertor in the Hamilton dynamics
$\delta_{\rm HD}$ and for the accept-reject step $\delta_{\rm acc}$
are fixed in accordance with \cite{GuKiSh}
\begin{equation}\label{tol_hd_acc}
\delta_{\rm HD}\propto 1/V, \qquad \delta_{\rm acc}\propto 1/V^2\,.
\end{equation}
As a consequence the average number of CG iterations can be estimated to
\begin{equation}\label{ncg}
\langle N^{\rm (CG)}\rangle \propto \zeta^{1/2}\ln V \,.
\end{equation}
The TSMB algorithm uses polynomials
within the least squares integral approximation of
the function $x^{-N_f/2}$ over the interval $[\epsilon,\lambda]$
(see \cite{Montv1,Montv2}).
The first (crude) polynomial of order $n_1$ is used in the
first multiboson step. The second (correcting) polynomial
of order $n_2$ is required for the noisy correction step.
The third polynomial of order $n_3$ approximates the inverse square
root of the second polynomial. And the fourth (finest)
polynomial of the order $n_4$ is necessary for the reweighting step.
These parameters are chosen according to a prescription in
\cite{HaMoMoOeScSk}:
\begin{eqnarray}
n_1 \propto (\sqrt{\zeta})^{1/2}\ln V, & & n_2 \simeq \langle%
N^{\rm (CG)}\rangle, \label{pol_ord_12} \\
n_3 \simeq (1.2 -\!\!\!- 1.4)n_2, & & n_4 \ge n_2,
\label{pol_ord_34} \\
\lambda \simeq (1.2 -\!\!\!- 1.4)\langle\lambda_{\rm max}\rangle,
& & \epsilon \simeq 0.5\langle\lambda_{\rm min}\rangle.
\label{intrv_marg}
\end{eqnarray}
For $n_1$ we have taken into account the empirical observation  
\cite{Montv2} that the dependence on the condition number
becomes effectively weaker for the least
squares first polynomial compared with polynomials in the  
Chebyshev approximation resulting in a replacement 
$\zeta\rightarrow\sqrt{\zeta}$.

Now let us come to the theoretical cost analysis. We define
the performance of an algorithm by
\begin{equation}\label{alg_prfrm}
{\bf P} = {\bf N}_{\rm oper}\tau_{\rm int},
\end{equation}
where ${\bf N}_{\rm oper}$ is the total number of operations per update.
$\tau_{\rm int}$ denotes the integrated autocorrelation time for a given
observable. The numbers of operations in the HMC and in the TSMB
($n_2\gg n_1$) algorithms are given as follows
\begin{equation}\label{nop}
{\bf N}^{\rm HMC}_{\rm oper}\propto V\langle N^{\rm (CG)}\rangle%
N_\tau, \qquad {\bf N}^{\rm TSMB}_{\rm oper}\propto V n_2.
\end{equation}
For simplicity instead of the value ${\bf N}_{\rm oper}$ we used 
the CPU time required for one sweep $t_{\rm CPU}$. 
The integrated autocorrelation times
for the HMC and the TSMB methods can be estimated 
according to arguments given in \cite{EdwHrvKe} and
\cite{Pear,Jans,Bor}
\begin{equation}\label{acrr}
\tau^{\rm HMC}_{\rm int}\propto(N_\tau\Delta\tau)^{-2}, \quad
\tau^{\rm TSMB}_{\rm int}\propto n_1 (\sqrt{\zeta})^{1/2}.
\end{equation}
Here the effective dependence on the condition number $\zeta$
(\ref{cond_num}) for the least squares first polynomial has been
taken into account, too. Thus, the expected theoretical gain ${\bf G}$ of
the TSMB algorithm over the HMC method is evaluated according to
(\ref{dt_nhmc}), (\ref{ncg}), (\ref{pol_ord_12}),
(\ref{alg_prfrm}--\ref{acrr}) as follows
\begin{equation}\label{gain_tsmb_hmc}
{\bf G}\equiv\frac{{\bf P}^{\rm HMC}}%
{{\bf P}^{\rm TSMB}}\propto\frac{V^{1/4}}{\ln V}.
\end{equation}
\begin{table}
\begin{center}
\begin{tabular}{|c|ccc|}
\hline
phase & $N_\tau$ & $\Delta\tau$ & $\langle N^{\rm (CG)}\rangle$ \\
\hline
Coulomb & 40 & 0.025 & 36.0(2) \\
confinement & 10 & 0.01 & 500(2) \\
\hline
\end{tabular}
\end{center}
\caption{HMC parameters for 
Coulomb and confinement phases. Stopping criteria are
$\delta_{\rm HD}=10^{-3}$ and $\delta_{\rm acc}=10^{-7}$.}
\label{tab:prm_hmc}
\vspace*{-0.1cm}
\end{table}

\begin{table}
\begin{center}
\begin{tabular}{|c|cccc|cc|}
\hline
 & $n_1$ & $n_2$ & $n_3$ & $n_4$ & $\epsilon$ & $\lambda$ \\
\hline
(a) & 6 & 36 & 48 & 200 & 0.025 & 2.5 \\
(b) & 50 & 360 & 450 & 500 & 0.000225 & 9 \\
\hline
\end{tabular}
\end{center}
\caption{TSMB parameters 
for Coulomb (a) and confinement (b) phases.}
\label{tab:prm_tsmb}
\vspace*{-0.1cm}
\end{table}

Now we turn to the numerical investigation of the performance of HMC
and TSMB algorithms both in the Coulomb and confinement phases.
The lattice size is $6^3\times 12$ and time-antiperiodic
boundary conditions for fermions are employed.
In the Coulomb phase we have chosen the point $(\beta, \kappa)=(2.0, 0.130)$, 
whereas in the  confinement phase $(\beta,\kappa)=(0.0, 0.238)$ 
quite close to the chiral limit.
For the HMC method the number of time steps $N_\tau$ and
the time step size $\Delta\tau$ selected according to the
prescription (\ref{dt_nhmc}) are presented in Table%
~\ref{tab:prm_hmc}. The stopping criteria for the CG method according to 
(\ref{tol_hd_acc}) provide an average number of CG
inversion steps (in the Hamiltonian dynamics)
$\langle N^{\rm (CG)}\rangle$ as shown in the same
Table. In case of the TSMB algorithm the polynomial parameters are
chosen according to the recipe (\ref{pol_ord_12}--\ref{intrv_marg})
and presented in Table~\ref{tab:prm_tsmb}. The knowledge of these
parameters requires information about
the smallest $\langle\lambda_{\rm min}\rangle$ and largest
$\langle\lambda_{\rm max}\rangle$ average eigenvalues of the even-odd
preconditioned fermion matrix ${\cal M}^\dag{\cal M}$, tunable to
the dynamical fermion case from the quenched one
(see Table~\ref{tab:egv_que_dyn}).

The performance of the dynamical fermion algorithms was studied with respect 
to the average values of the following gauge invariant observables
${\cal O}$: the mean gauge plaquette energy
$\langle E_G\rangle = \langle 1- \,{\rm Re}\,U_{\rm p} \rangle$ 
the scalar condensate
$\langle \overline{\psi}\psi \rangle$ and the pion norm
$\langle \Pi \rangle = \langle(\overline{\psi}\gamma_5\psi)^2\rangle $. 
We measured the
corresponding integrated autocorrelation times $\tau_{\rm int}$
and also the gain ${\bf G}$ (\ref{gain_tsmb_hmc})
of the TSMB over the HMC method.

\begin{table}
\begin{center}
\begin{tabular}{|c|cc|cc|}
\hline
 & \multicolumn{2}{c}{quenched} \vline &
 \multicolumn{2}{c}{dynamical} \vline \\
\hline
 & $\langle\lambda_{\rm min}\rangle$
 & $\langle\lambda_{\rm max}\rangle$
 & $\langle\lambda_{\rm min}\rangle$
 & $\langle\lambda_{\rm max}\rangle$ \\ 
\hline
(a) & .065(1) & 1.60(1) & .13(1) & 1.63(1) \\ 
(b) & .0010(1) & 6.78(1) & .0005(1) & 6.59(1) \\
\hline
\end{tabular}
\end{center}
\caption{Minimal $\langle\lambda_{\rm min}\rangle$  and maximal
$\langle\lambda_{\rm max}\rangle$ average eigenvalue of the even-odd
preconditioned Wilson fermion matrix ${\cal M}^\dag{\cal M}$ for
both Coulomb (a) and confinement (b) phases in quenched and dynamical
cases, the latter determined from HMC.}
\label{tab:egv_que_dyn}
\vspace*{-0.1cm}
\end{table}

\begin{table}
\begin{center}
\begin{tabular}{|c|ccc|}
\hline
 & $\langle E_G\rangle$ & $\langle\overline{\psi}%
\psi\rangle$ & $\langle\Pi\rangle$ \\
\hline
& \multicolumn{3}{c}{Coulomb} \vline
\\
\hline
$\langle{\cal O}^{\rm HMC}\rangle$ & 0.1332(1) &
0.9381(1) & 1.378(1) \\
$\langle{\cal O}^{\rm TSMB}\rangle$ & 0.1331(1) &
0.9379(1) & 1.376(1) \\
\hline
$\tau^{\rm HMC}_{\rm int}$ & 3.2(3) & 2.0(2) & 25(4) \\
$\tau^{\rm TSMB}_{\rm int}$ & 3.0(3) & 2.8(2) & 50(8) \\
\hline
${\bf G}$ & 1.7(2) & 1.2(2) & 0.8(2) \\
\hline
& \multicolumn{3}{c}{Confinement} \vline
\\
\hline
$\langle{\cal O}^{\rm HMC}\rangle$ &
0.939(1) & 0.95(1) & 13.9(2) \\
$\langle{\cal O}^{\rm TSMB}\rangle$ &
0.938(1) & 0.96(1) & 13.7(2) \\
\hline
$\tau^{\rm HMC}_{\rm int}$ & 65(7) & 60(7) & 35(5) \\
$\tau^{\rm TSMB}_{\rm int}$ & 120(20) & 125(15) & 45(5) \\
\hline
${\bf G}$ & 0.5(1) & 0.5(1) & 0.7(1) \\
\hline
\end{tabular}
\end{center}
\caption{Performance of HMC and TSMB algorithms in the
Coulomb and confinement phases.}
\label{perf_hmc_tsmb}
\vspace*{-0.1cm}
\end{table}

The main result of our investigation is presented in
Table~\ref{perf_hmc_tsmb}. The statistics in our case
($O(10^4)$ measurements) was sufficient to evaluate the integrated
autocorrelation time. The reweighting procedure in the TSMB
case has not changed significantly the results obtained
using only the main two steps. 
We may conclude that the performance of the two algorithms is
comparable for the chosen parameters. Further optimization of
TSMB is possible by increasing the number of local gauge updates from one
(in our case) to a larger number proportional to
$\tau^{\rm TSMB}_{\rm int}$ in (\ref{acrr}).
Moreover, the polynomial degrees $n_1$ and $n_2$ could be
taken smaller leaving more space for error correction by
reweighting. This shows that for an even number of fermion flavours
the TSMB algorithm is at least competitive with HMC.

This research has been supported by the graduate college DFG-GK 271, 
by the RFRB grant 99-01-01230 and by the JINR Heisenberg-Landau program.


\begin{thebibliography}{99}
%
\bibitem{HfMiMPNhSt} A. Hoferichter, V. Mitrjushkin,
                     M. M\"uller-Preussker, T. Neuhaus and
                     H. St\"uben, Nucl. Phys. B434 (1995) 358.

\bibitem{DuKePeRow}  S. Duane, A. Kennedy, B. Pendleton and
                     D. Roweth, Phys. Lett. B195 (1987) 216.

\bibitem{HfMiMPSt}  A. Hoferichter, V. Mitrjushkin,
                    M. M\"uller-Preussker and H. St\"uben,
                    Phys. Rev. D58 (1998) 114505.

\bibitem{Montv1}  I. Montvay, Nucl. Phys. B466 (1996) 259.

\bibitem{Lusch}  M. L\"uscher, Nucl. Phys. B418 (1994) 637.

\bibitem{Pear}  M. Peardon (UKQCD Collaboration), Nucl. Phys.
                (Proc. Suppl.) B42 (1995) 891.
                
\bibitem{BorFor} A. Borici and Ph. de Forcrand, 
                Nucl. Phys. B454 (1995) 645.

\bibitem{Jans}  K. Jansen, Nucl. Phys. (Proc. Suppl.) B53
                (1997) 127.

\bibitem{Bor}  A. Borici, hep-lat/9602018.

\bibitem{EdwHrvKe}  R. Edwards, I. Horvath and A. Kennedy,
                    Nucl. Phys. B484 (1997) 375.

\bibitem{GuKiSh}  R. Gupta, G. Kilcup and S. Sharpe,
                  Phys. Rev. D38 (1988) 1278.

\bibitem{Montv2}  I. Montvay, Comput. Phys. Commun. 109
                  (1998) 144.

\bibitem{HaMoMoOeScSk}  S. Hands, I. Montvay, S. Morrison,
                        M. Oevers, L. Scorzato and J. Skullerud,
                        Eur. Phys. J. C17 (2000) 285.

\end{thebibliography}
\end{document}